\documentclass[usenatbib]{mn2e}

\usepackage{graphicx}
\usepackage{natbib}
\usepackage{multirow}
\usepackage{fixltx2e}
\usepackage{float}
\usepackage{subfloat}
\def\subfloatcap{(\alph{subfloatnumber})}

\def\citeN{\citet}
\def\cite{\citep}

\footnotesize
\newdimen\digitwidth    
\setbox0=\hbox{\rm0}
\digitwidth=\wd0
\catcode`!=\active
\def!{\kern\digitwidth}
\normalsize
\title[Polarisation measurements of five pulsars]{Polarisation measurements of five pulsars with interpulses}
\author[M.~J.~Keith et al.]
{M.~J.~Keith$^{1}$\thanks{Email: mkeith@pulsarastronomy.net},
S.~Johnston$^{1}$,
P.~Weltevrede$^{1}$ and
M.~Kramer$^{2,3}$
\\
$^1$ Australia Telescope National Facility, CSIRO, P.O. Box 76, Epping, NSW 1710, Australia\\
$^2$ Max-Planck-Institut f\"ur Radioastronomie, Auf dem Huegel 69, 53121 Bonn, Germany\\
$^3$ University of Manchester, Jodrell Bank Centre for Astrophysics, Alan Turing Building, Manchester M13 9PL, UK\\
}
\let\leqslant=\leq 
\date{}
\begin{document}

\maketitle
\newcommand{\setthebls}{
}

\setthebls

\begin{abstract} 
We present polarisation observations of five pulsars
whose profiles exhibit two distinct emission regions separated
by close to 180\degr\ of longitude.
We fitted the position angle of the linear polarisation 
using the rotating vector model and convincingly show that all the pulsars
have the angle between their magnetic and 
rotation axes close to 90\degr. The simplest interpretation of the
results is that we see `main pulse' emission from one pole and `interpulse'
emission from the opposite pole.
We have attempted to produce emission maps of the magnetosphere above
the polar caps for each pulsar and
find that the maps support the view that the emission region in pulsars
is complex, even when the profile appears simple.
For three pulsars, we can derive
emission heights and polar maps which are consistent with emission regions
located symmetrically about the magnetic axis and confined
to the open field lines. For two pulsars, we find that either the emission
arises from `closed' field lines or that the profiles are highly asymmetric
with respect to the magnetic axis.

\end{abstract}


\section{Introduction}
The measured width of pulsar profiles is typically less than 10 percent
of the pulse period.  Exceptions to this rule exist and can
broadly be classified into three types \cite{wj08}.
In the first type, which we
refer to as interpulse pulsars, the profile has two distinct regions of
emissions separated by close to 180\degr. The interpretation is that
we see emission from open field lines near the two magnetic poles of the
pulsar when the magnetic and rotation axes are orthogonal.
In the second type, emission
occurs over a large fraction of the pulse period; these are generally
thought to be pulsars in which the magnetic and rotation axes are
close to alignment. Finally, in younger pulsars, wide double profiles
are observed the interpretation of which is that either the height from
which the radio emission originates is rather large \cite{jw06,kj07}
or the emission arises in a fan-beam \cite{man96}.

An alternative physical model for the interpulse pulsars suggests that emission 
from both the main pulse (MP) and interpulse (IP) actually originate 
from the same magnetic pole \cite{dzg05}.  In this model, the MP is seen as 
normal emission as the axis crosses close to the line of sight, and the 
IP is seen as `inward' emission from the same pole as it crosses 
directly opposite to the line of sight.

For this paper, we carried out polarisation observations of
a selection of five pulsars with strong IP emission, and applied the 
rotating vector model (RVM; \citealp{rc69a})
to determine their geometry.
PSR~J0627+0706 was discovered in the Perseus Arm 
pulsar survey (as yet unpublished).  The other four pulsars,
PSRs J1549--4848, J1722--3712, J1739--2903 and J1828--1101, are 
previously discovered \cite{mld+96,mlt+78,cl86,mhl+02}, but have little or 
no published polarisation data.
We describe the observations and the method used to fit RVM in
Section \ref{obs_sec} and
discuss how to construct polar maps in Section \ref{sec_polarmaps}.
In Section \ref{pulsars} we present polarisation profiles and derive 
emission heights for each pulsar in turn.
In Section \ref{discussion} we discuss the implications of these results
and provide a short conclusion in Section \ref{concs}.

\section{Observations and RVM Fits}
\label{obs_sec}
\label{rvm_sec}
\label{errors}
The five pulsars were observed using the Parkes 64-m radio telescope 
with data recorded using a digital spectrometer.
Each pulsar was observed at 1.4~GHz
with an integration time sufficient to yield a high signal-to-noise
ratio. PSR~J1828--1101 is heavily scattered at 1.4~GHz; we
therefore observed it at 3.1~GHz.
For details of the observing setup and calibration procedure, readers 
should refer to \citeN{jkmg08} and \citeN{wj08}.

\begin{table*}
\caption{
\label{rvm_table}
Parameters for seven pulsars from fits to the RVM.
}
\begin{tabular}{lr@{ $\pm$ }lr@{ $\pm$ }lr@{ $\pm$ }lr@{ $\pm$ }lr@{ $\pm$ }lcr@{ $\pm$ }lc}
Name            &\multicolumn{2}{c}{$\alpha_{\rm MP}$} & \multicolumn{2}{c}{$\beta_{\rm MP}$}  & \multicolumn{2}{c}{$\alpha_{\rm IP}$}  & \multicolumn{2}{c}{$\beta_{\rm IP}$}  & \multicolumn{2}{c}{$\phi_0$} & $\Delta\phi_{\rm I}$ & \multicolumn{2}{c}{$\Delta\phi_{\rm PA}$}& OG\\
\hline
J0627+0706  & 86.0 & 0.2    & 8.7   & 0.3    & 94.0  & 0.3    & 0.7  & 0.1 & 94.8  & 0.2   & 176.2 & 179.4&0.4  & MP\\
J0908--4913 & 96.1 & 0.4    & -5.9  & 0.6    & 83.9  & 0.2    & 6.3  & 0.4 & 96.42 & 0.07  & 184.5 &179.99&0.06 & IP \\
B1055--52   & 75.2 & 0.4    & 36.1  & 0.6    & 104.8 & 0.4    & 6.1  & 0.6 & 137.4 & 0.6   & 204   &189.5 &1.3  & MP \\
J1549--4848 & 92.5 & 0.2    & -3.5  & 0.2    & 87.5  & 0.2    & 1.5  & 0.2 & 93.0  & 0.2   & 180   &183.4 &0.3  & MP \\ 
J1722--3712 & 90.7 & 0.1    & 5.4   & 0.3    & 89.3  & 0.1    & 6.7  & 0.5 & 95.0  & 0.1   & 179.7 &181.5 &1.2  & IP\\
J1739--2903 & 84.2 & 0.3    & 3.3   & 0.2    & 95.8  & 0.2    & -8.2 & 0.4 & 88.1  & 0.1   & 180.7 &182.7 &0.7  & IP \\
J1828--1101 & 97.31 & 0.6   & -10.6 & 1.5    & 82.7  & 0.6    & 4.0  & 1.4 & 94.8  & 0.7   & 180.2 &178   &6    & MP \\
\end{tabular}
\end{table*}

\begin{table*}
\caption{
\label{der_table}
Parameters for seven pulsars used to derive their polar maps.
}
\begin{tabular}{lcccccccc}
Name & Period & \multicolumn{2}{c}{MP} & \multicolumn{2}{c}{IP} & $R_{\rm LC}$ &  $r_{\rm em}$ & $\Delta \phi$ \\
& (s) & ($\phi_{\rm l} - \phi_0$) & ($\phi_{\rm t} - \phi_0$) & ($\phi_{\rm l} - \phi_0$) & ($\phi_{\rm t} - \phi_0$) & (km) & (km) & (deg) \\ \hline
J0627+0706  & 0.47588 & --6.6 & 1.4 & --10.0& 6.9 & 22700 & 160/320 & 1.6/3.2 \\
J0908--4913 & 0.10675 & --16.5& 3.5 & --19  & 0   & 5100  & 245     & 11      \\
B1055--52   & 0.19711 & --15  & 22  & --42  & 10  & 9400  & 730     & 18      \\
J1549--4848 & 0.28835 & --11.5& 7   & --11.5& 4   & 13700 & 180     & 2.9     \\
J1722--3712 & 0.23617 & --17  & 5   & --13.5& 1   & 11300 & 290     & 6.0     \\
J1739--2903 & 0.32288 & --8.1 & 9.9 & --9.1 & 9.9 & 15400 & 45/230  & 0.7/3.4 \\
J1828--1101 & 0.07205 & --11.3& 4.2 & --10.8& 3.2 & 3400  & 55      & 3.5     \\
\end{tabular}
\end{table*}

The RVM shows that the characteristic S-shaped swing of the 
polarisation position angle (PA) can be explained if the PA of the
linear polarisation (LP) is defined by the direction of the magnetic 
field at the point of emission \cite{rc69a}.
If one assumes that the polarised emission is well modelled by the RVM, then one can use polarisation measurements to constrain the geometry.
The geometry of a pulsar is characterised by two parameters, $\alpha$, 
the angle between the magnetic and rotation axes, and $\beta$, 
the minimum angle between the magnetic axis and the line of 
sight to the observer. The RVM also allows derivation of two other 
parameters, $\phi_0$ which is the pulse longitude at which the PA swing 
has the steepest gradient, and $\Psi_0$ which is the PA at that longitude.
In this paper we adopt the convention that the angle between the rotation axis and the line of sight to the observer, $\zeta = \alpha + \beta$.
The RVM states that the observed PA, $\Psi$, depends on the pulse longitude, $\phi$, and the geometric parameters as
\begin{equation}
\label{equ:rvm}
\tan (\Psi+\Psi_0) = \frac{ \sin \alpha \; \sin(\phi-\phi_0)}{
\sin\zeta\; \cos\alpha - \cos\zeta\sin\alpha
\cos(\phi-\phi_0)}.
\end{equation}
We follow the `RVM sign convention' for the PA (see \citealp{ew01}).

For the majority of pulsars, the range of pulse longitude for which the 
PA can be measured (i.e. the range where there is significant emission)
is small.  When this is the case, the RVM fit for $\alpha$ and 
$\beta$ is highly degenerate and it is difficult to constrain the geometry.
However, in the case of pulsars with an IP, we can usually measure the 
PA in two, well-separated, regions of pulse longitude and therefore 
break this degeneracy. 
Although \citeN{wj08b} tabulated a list of 27 pulsars, which they considered
to be an upper limit to the number of orthogonal rotators in the observed
population, only the recent result on PSR~J0908--4913 by \citeN{kj08}
shows incontrovertible evidence for this directly from RVM fits.
This is partly because the PA swings of many pulsars deviate from 
purely geometrical values due to effects both in the pulsar 
magnetosphere (e.g. orthogonal mode emission; \citealp{kkj+02}) and in the 
interstellar medium (e.g. scattering effects; \citealp{kar09}).

The RVM was fitted to the PA swing for the five pulsars in our sample 
using a non-linear ordinary least-squares algorithm.
Table \ref{rvm_table} gives the results for the five pulsars 
and for PSRs J0908--4913 and B1055--52 for which high signal-to-noise
data have been presented elsewhere \cite{kj08, ww09}.
In addition to the geometric parameters, we 
also present the separation of the steepest gradient in
the MP and IP, $\Delta\phi_{\rm PA}$.
This is computed by fitting independent values of $\phi_0$ for the 
PAs of the MP and IP.
All five parameters are then fitted simultaneously.

The confidence contours shown in later figures and the standard errors in
Table \ref{rvm_table} have been generated by scaling the observed PA errors 
so that $\chi^2 = 1$.  These errors assume that the RVM model 
can fully describe the observed PA swing. 
It is clear however that the residuals have small but significant unmodeled variability, and the errors we present are likely to underestimate the true error in the model parameters.

\subsection{Implications for gamma-ray emission}
The {\it Fermi} satellite has recently detected a large number of
gamma-ray pulsars \cite{abdo9}.
The implication of the detections is
that gamma-ray emission arises in the outer magnetosphere. In the
so-called `outer-gap' model \cite{chr86a,rom96a}, gamma-rays are only 
observed from field lines above the null charge line. For pulsars in
which both radio and gamma-ray emission are detected, the gamma-ray 
emission is associated with the opposite magnetic pole to the radio emission.
However, in the case of orthogonal rotators in which radio emission is seen 
from both poles, knowledge of $\alpha$ and $\beta$ allows us to
determine whether the MP or the IP emission originates from above the same
pole as any putative gamma-ray emission.
This is listed in the final column of Table~\ref{rvm_table}. Of our sample,
only PSR~B1055--52 has been detected in gamma-rays to date but on-going
{\it Fermi} observations may uncover further detections.

\begin{figure*}
\begin{minipage}{.49\textwidth}
\center
\includegraphics[width=7.5cm]{0627_z}
\subcaption*{}
\label{0627_z}
\end{minipage}
\begin{minipage}[t]{.49\textwidth}
\begin{minipage}[b]{.99\textwidth}
\begin{minipage}[b]{.45\textwidth}
\center
\includegraphics[height=4.2cm,angle=-90]{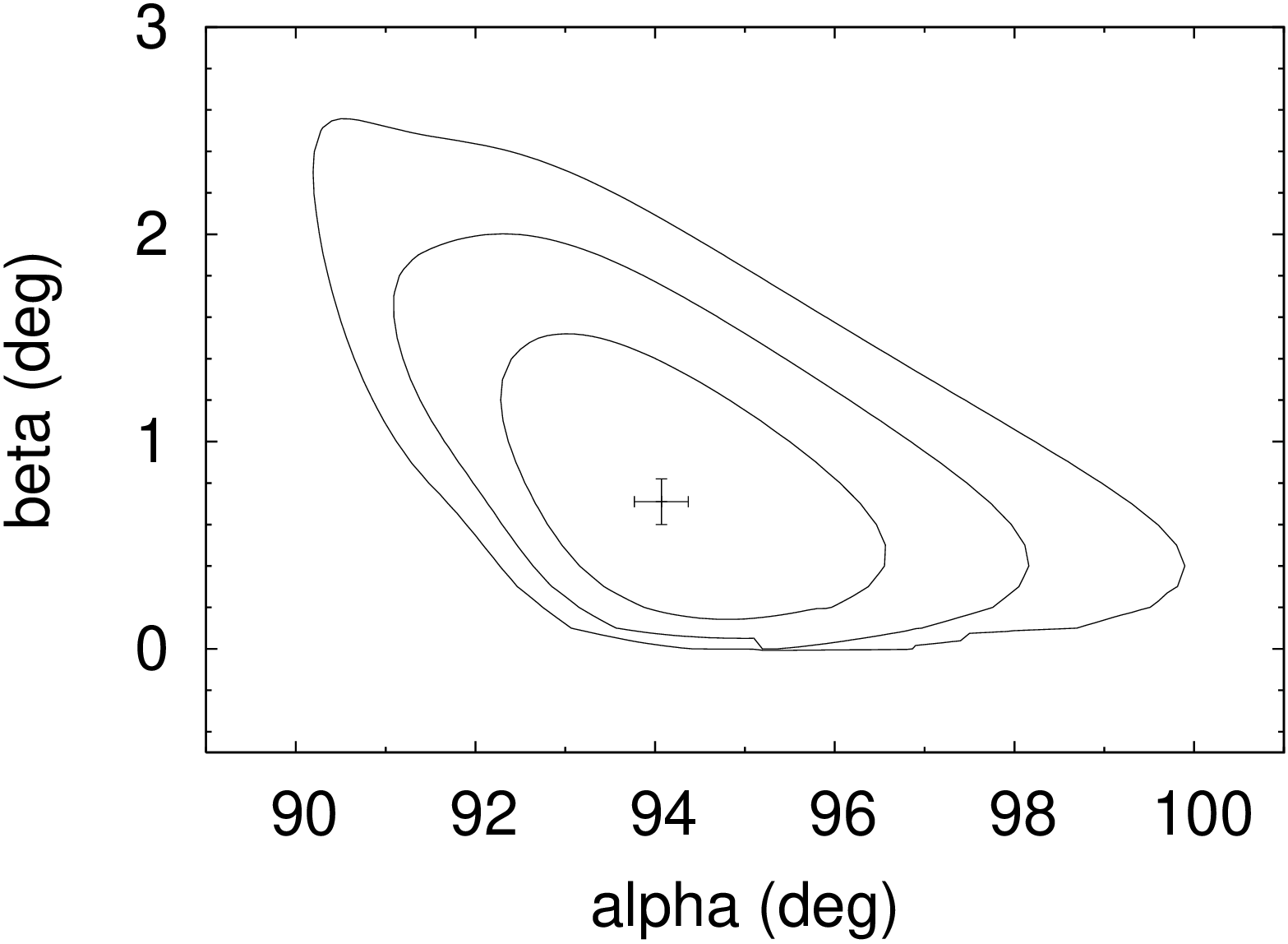}
\subcaption*{}
\label{0627_cont}
\end{minipage}
\begin{minipage}[b]{.45\textwidth}
\center
\includegraphics[height=5.5cm]{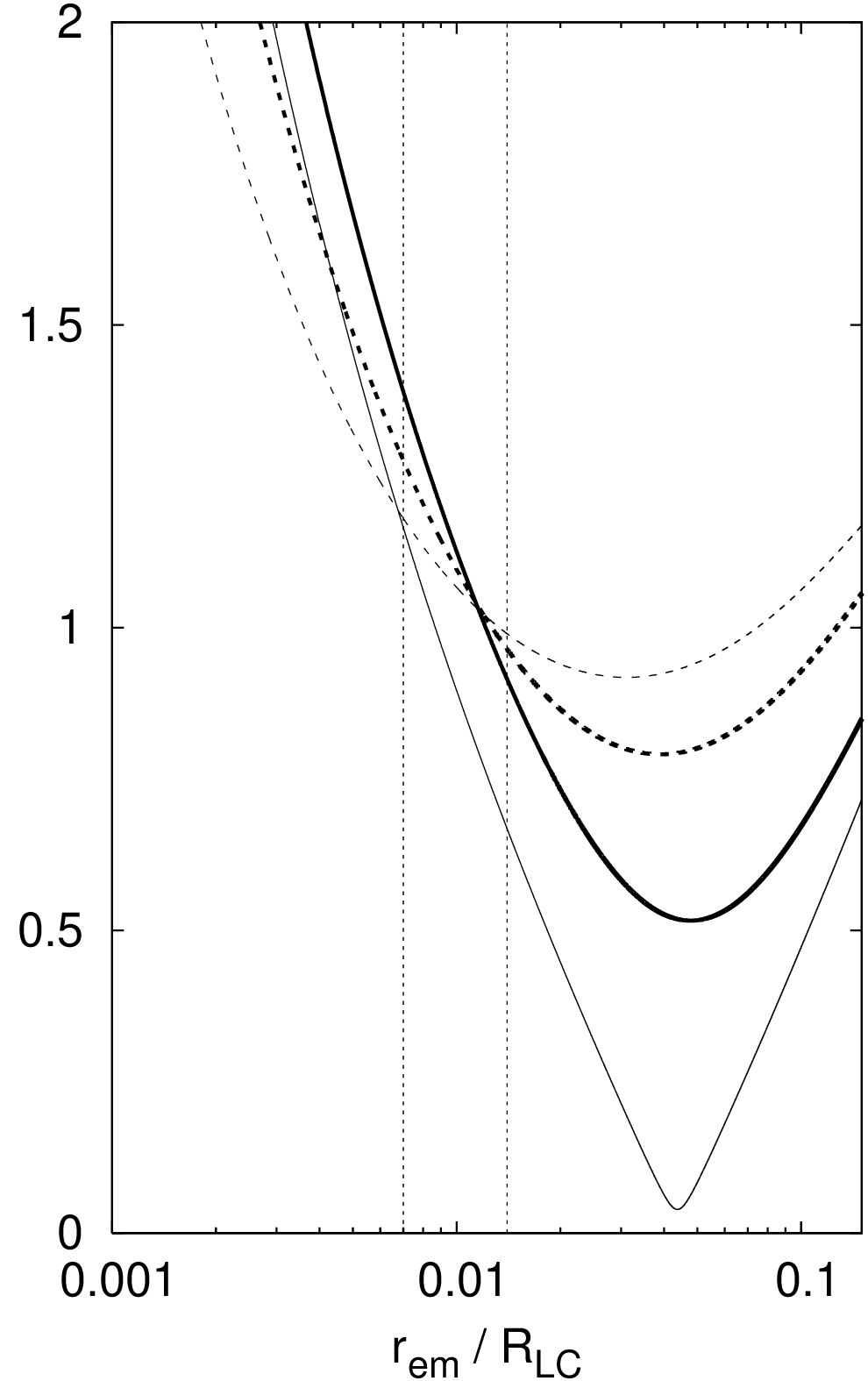}
\subcaption*{}
\label{0627_splot}
\end{minipage}
\end{minipage}
\begin{minipage}[b]{.99\textwidth}
\begin{minipage}[t]{.45\textwidth}
\center
\includegraphics[height=3.8cm,angle=-90]{0627_beammap1}
\subcaption*{}
\label{0627_beammap1}
\end{minipage}
\begin{minipage}[t]{.45\textwidth}
\center
\includegraphics[height=3.8cm,angle=-90]{0627_beammap2}
\subcaption*{}
\label{0627_beammap2}
\end{minipage}
\end{minipage}
\end{minipage}
\caption[]{
\label{0627_figs}
(a) The 1.4~GHz integrated profile of PSR~J0627+0706 near $\phi=\pm90$\degr.
The peak flux density has been placed at a longitude of 90\degr.  The lower 
part of the figure shows the total intensity (black), LP (red) and CP (blue).
The upper part of the figure shows the measured PA and the RVM fit.
An orthogonal jump can be seen near phase $-$88\degr.
(b)  Contours of constant $\chi^2$ in the residuals of the RVM fit 
as a function of $\alpha$ and $\beta$.
The contours approximate 1,2 and 3-$\sigma$ confidence limits.
The cross marks the position of the minimum $\chi^2$.
(c) $s$ values for the leading (solid lines) and trailing (dashed lines) edge 
of the MP (thick lines) and IP (thin lines) as a function of $r_{\rm em}$.
The vertical lines show choices of $r_{\rm em}$ used for Figures (d) and (e).
(d) and (e) polar cap emission maps for $r_{\rm em} = 160$ 
and 320~km respectively.
The MP and IP have been mapped onto the geometric model, with the IP
reflected horizontally so that field lines are physically identical 
for both MP and IP.  The inner circle denotes the extent of the open field 
line region, $s=1$, and the outer circle is twice this extent.
}
\end{figure*}

\begin{figure*}
\begin{minipage}{.49\textwidth}
\center
\includegraphics[width=7.5cm]{1549_z}
\subcaption*{}
\label{1549_z}
\end{minipage}
\begin{minipage}[t]{.49\textwidth}
\begin{minipage}[b]{.99\textwidth}
\begin{minipage}[b]{.45\textwidth}
\center
\includegraphics[height=4.2cm,angle=-90]{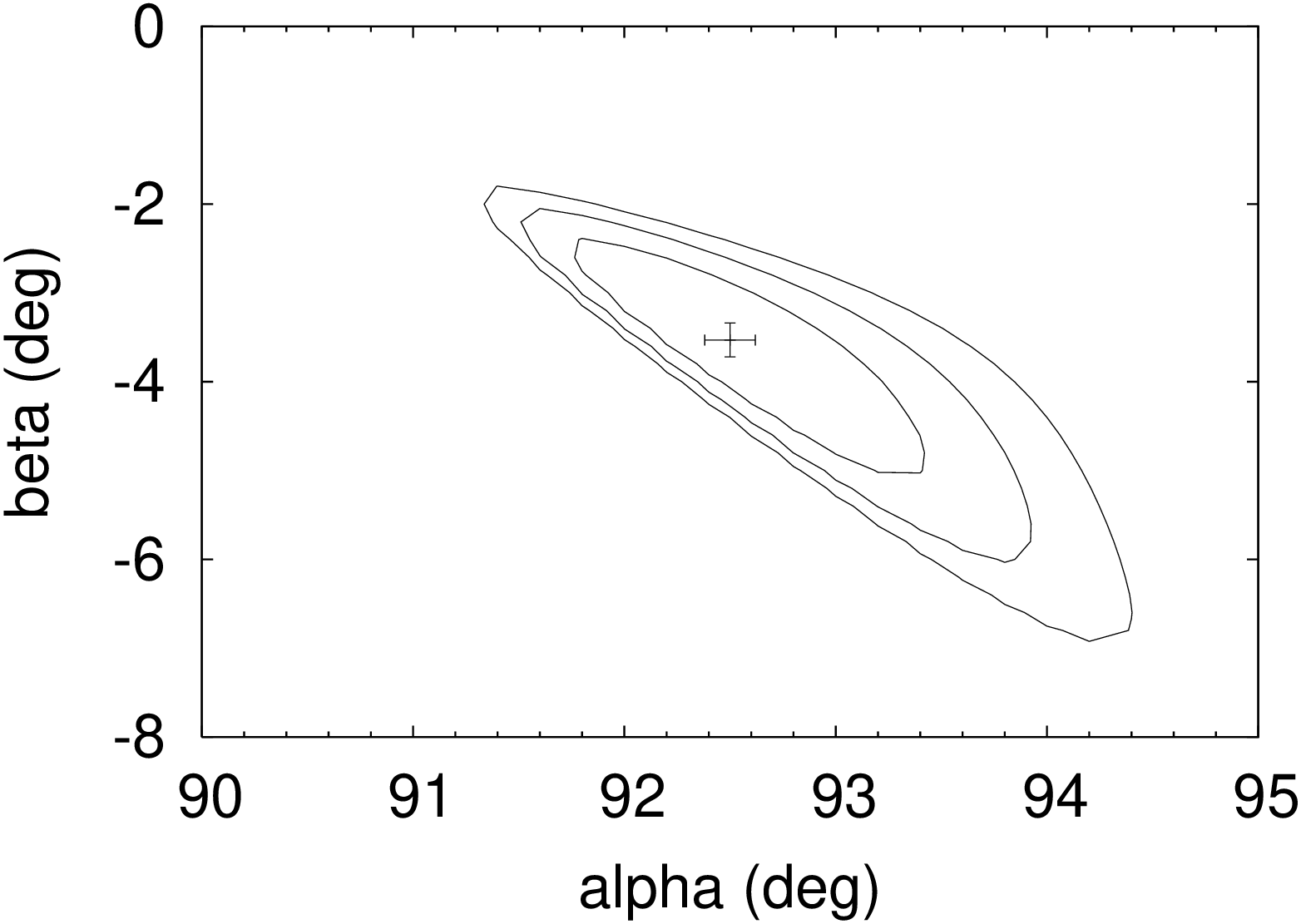}
\subcaption*{}
\label{1549_cont}
\end{minipage}
\begin{minipage}[b]{.45\textwidth}
\center
\includegraphics[height=5.5cm]{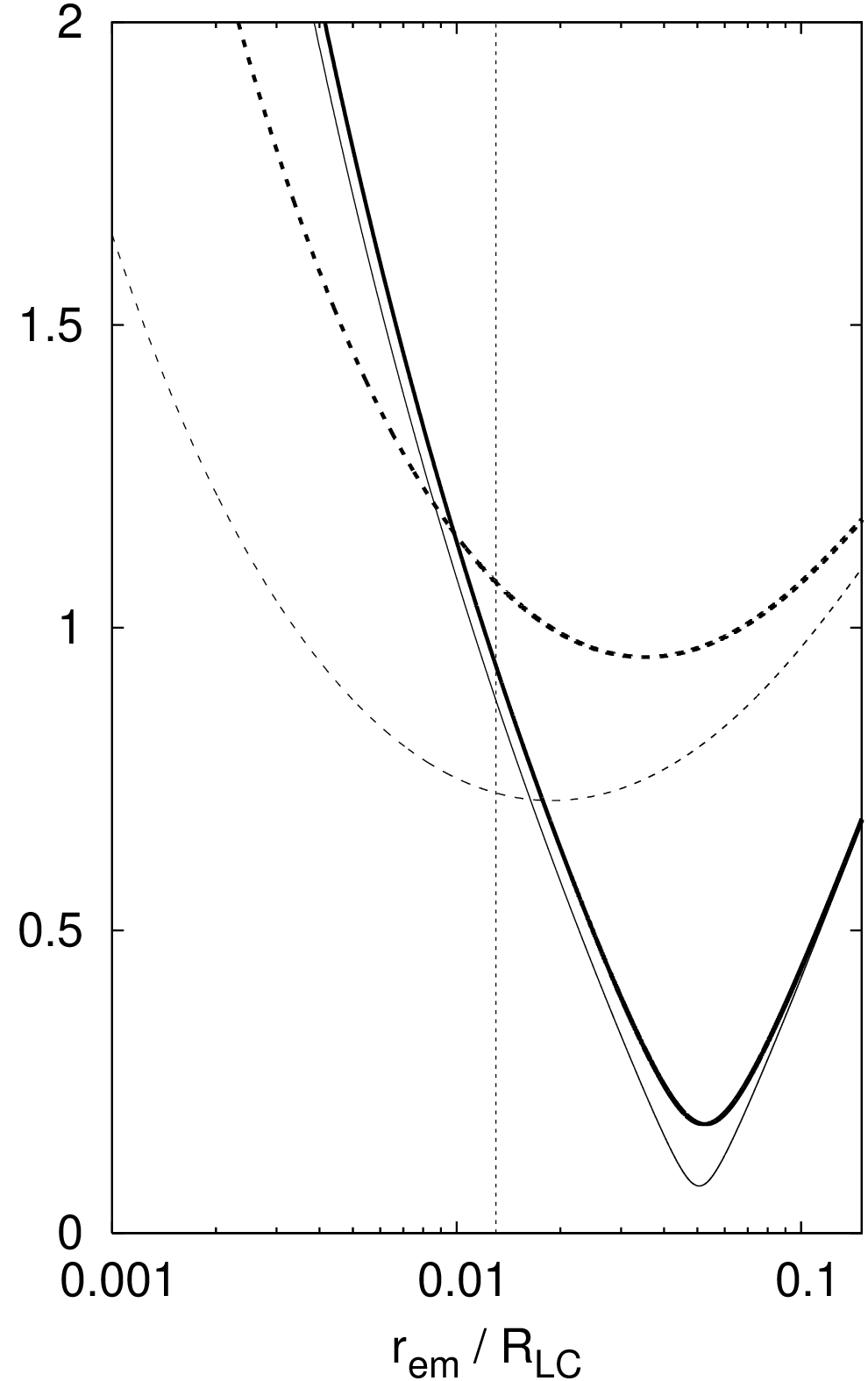}
\subcaption*{}
\label{1549_splot}
\end{minipage}
\end{minipage}
\begin{minipage}[b]{.99\textwidth}
\center
\includegraphics[height=3.8cm,angle=-90]{1549_beammap}
\subcaption*{}
\label{1549_beammap}
\end{minipage}
\end{minipage}
\caption[]{
\label{1549_figs}
(a) The 1.4~GHz integrated pulse profile of PSR~J1549--4848, clipped to show the details of the two pulses.
An orthogonal jump can be seen near phase of $-$87\degr.
The format of this plot, and the others in this figure are the same as for Figure \ref{0627_figs}.
The polar cap emission map in (d) assumes $r_{\rm em}$ = 180~km.
}
\end{figure*}
\begin{figure*}
\begin{minipage}{.49\textwidth}
\center
\includegraphics[width=7.5cm]{1722_z}
\subcaption*{}
\label{1722_z}
\end{minipage}
\begin{minipage}[t]{.49\textwidth}
\begin{minipage}[b]{.99\textwidth}
\begin{minipage}[b]{.45\textwidth}
\center
\includegraphics[height=4.2cm,angle=-90]{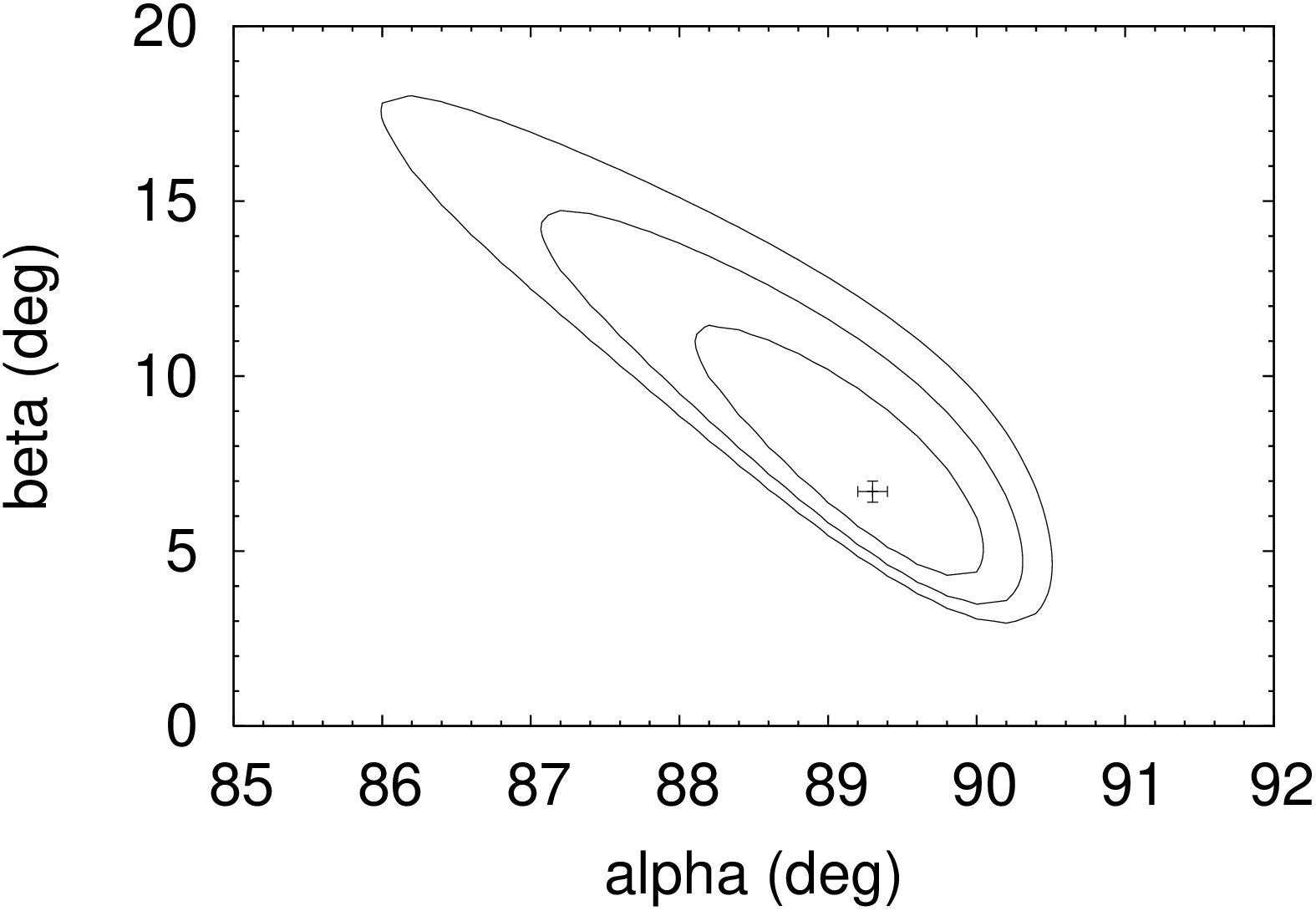}
\subcaption*{}
\label{1722_cont}
\end{minipage}
\begin{minipage}[b]{.45\textwidth}
\center
\includegraphics[height=5.5cm]{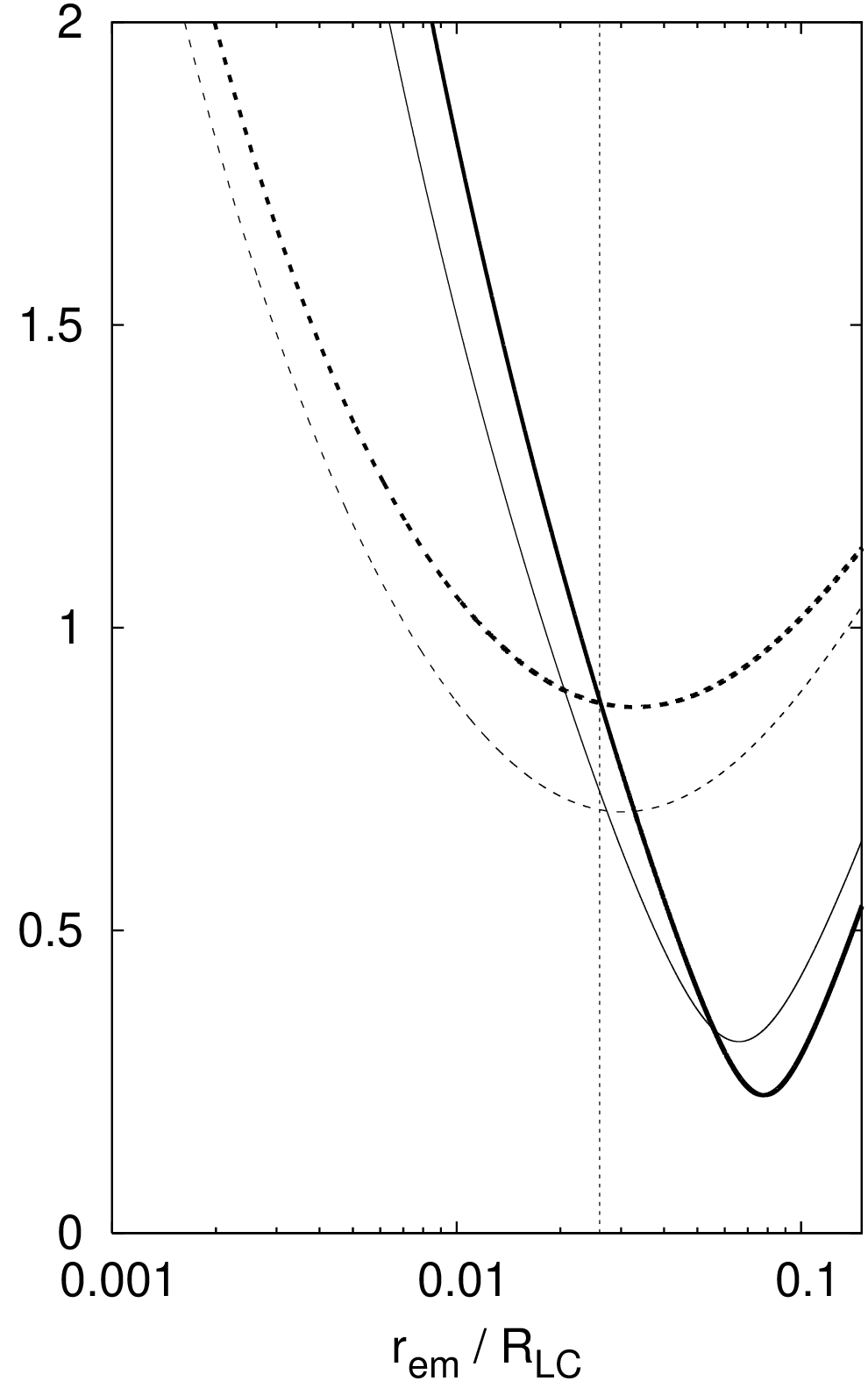}
\subcaption*{}
\label{1722_splot}
\end{minipage}
\end{minipage}
\begin{minipage}[b]{.99\textwidth}
\center
\includegraphics[height=3.8cm,angle=-90]{1722_beammap}
\subcaption*{}
\label{1722_beammap}
\end{minipage}
\end{minipage}

\caption[]{
\label{1722_figs}
(a) The 1.4~GHz integrated pulse profile of PSR~J1722--3712, clipped to show the details of the two pulses.
The format of this plot, and the others in this figure are the same as for Figure \ref{0627_figs}.
The polar cap emission map in (d) assumes $r_{\rm em}$ = 290~km.
}
\end{figure*}

\begin{figure*}

\begin{minipage}{.49\textwidth}
\center
\includegraphics[width=7.5cm]{1739_z}
\subcaption*{}
\label{1739_z}
\end{minipage}
\begin{minipage}[t]{.49\textwidth}
\begin{minipage}[b]{.99\textwidth}
\begin{minipage}[b]{.45\textwidth}
\center
\includegraphics[height=4.2cm,angle=-90]{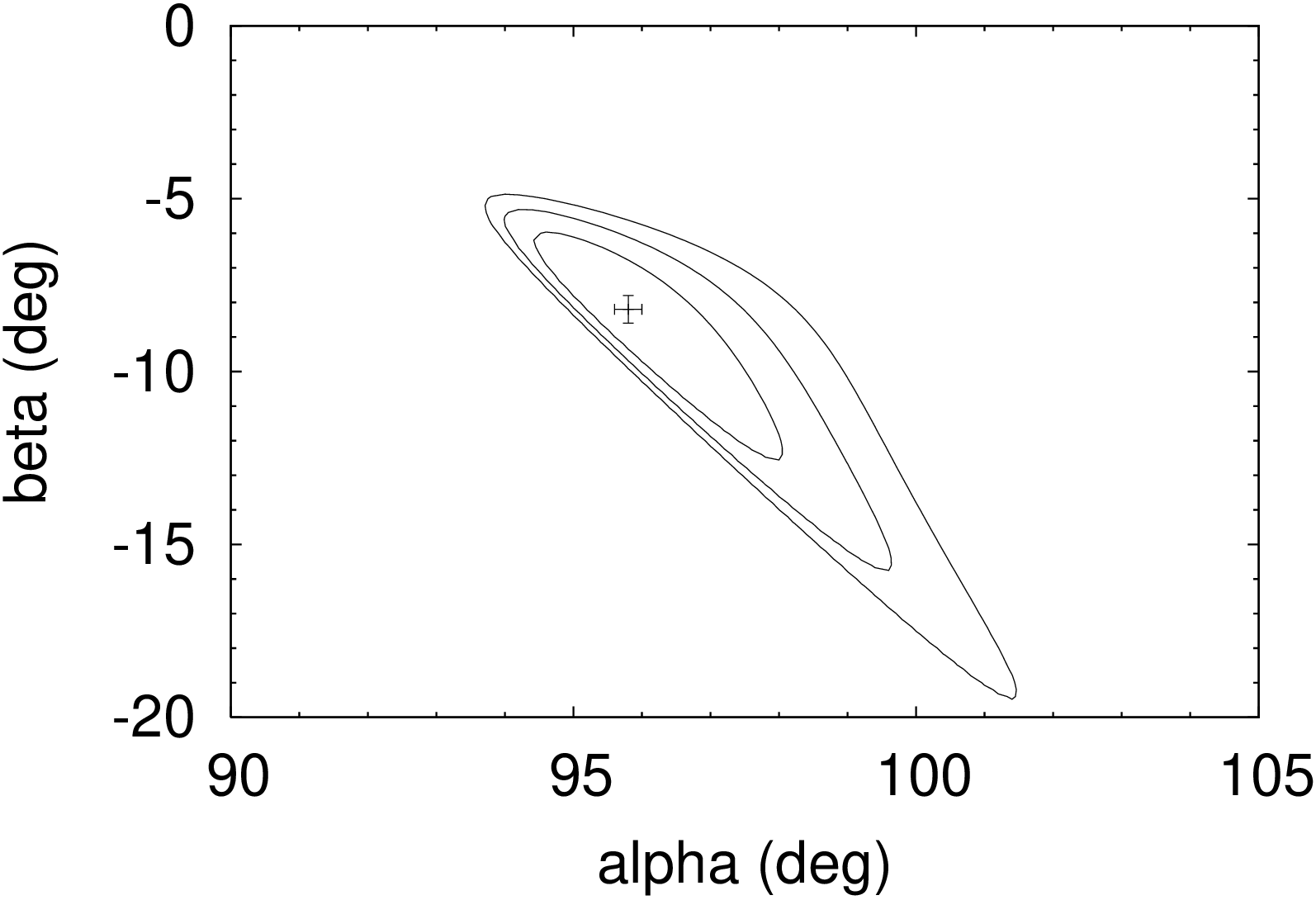}
\subcaption*{}
\label{1739_cont}
\end{minipage}
\begin{minipage}[b]{.45\textwidth}
\center
\includegraphics[height=5.5cm]{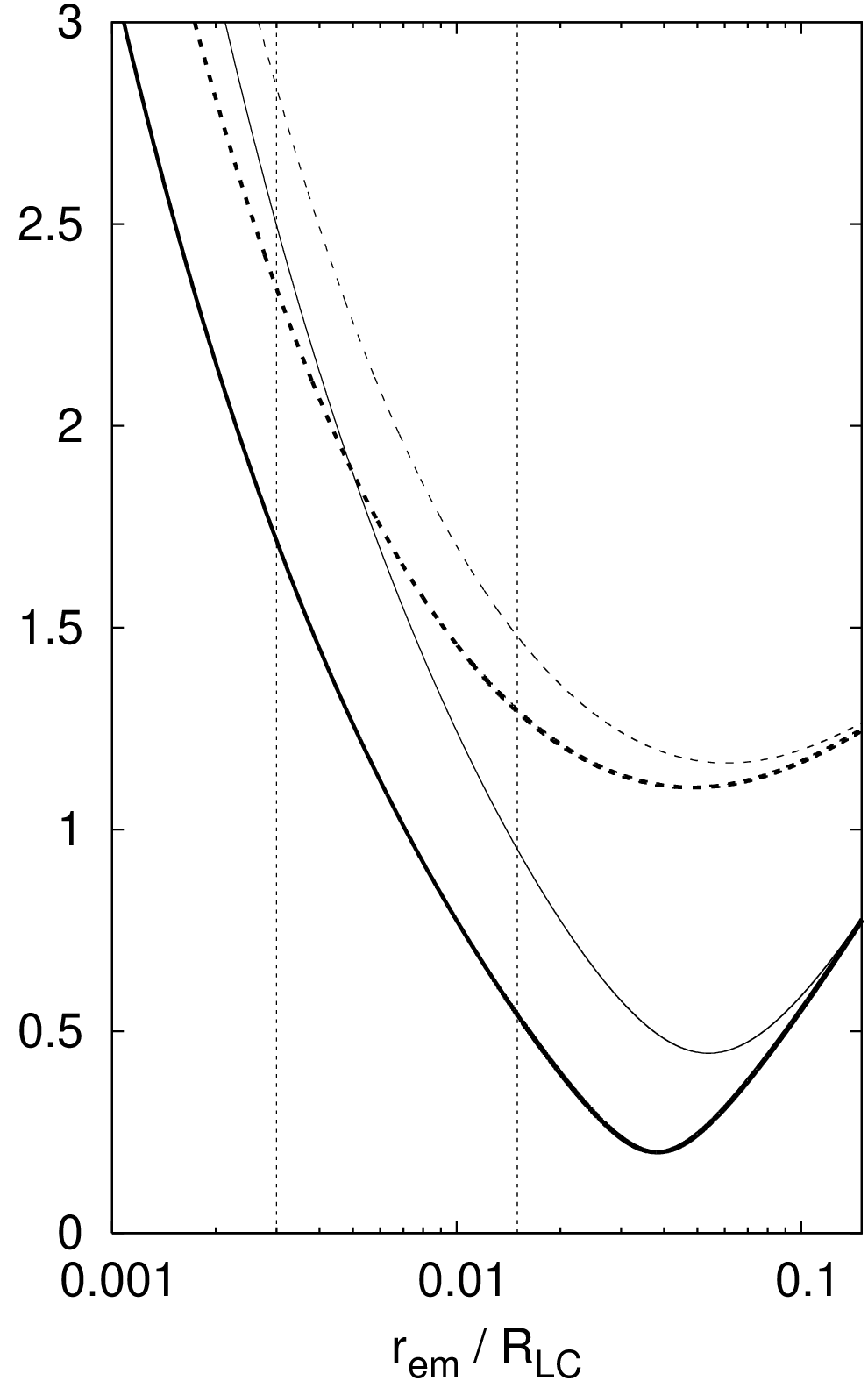}
\subcaption*{}
\label{1739_splot}
\end{minipage}
\end{minipage}
\begin{minipage}[b]{.99\textwidth}
\begin{minipage}[t]{.45\textwidth}
\center
\includegraphics[height=3.8cm,angle=-90]{1739_beammap1}
\subcaption*{}
\label{1739_beammap1}
\end{minipage}
\begin{minipage}[t]{.45\textwidth}
\center
\includegraphics[height=3.8cm,angle=-90]{1739_beammap2}
\subcaption*{}
\label{1739_beammap2}
\end{minipage}
\end{minipage}
\end{minipage}
\caption[]{
\label{1739_figs}
(a) The 1.4~GHz integrated pulse profile of PSR~J1739--2903, clipped to show the details of the two pulses.
An orthogonal jump can be seen around a phase of 89\degr.
The format of this plot, and the others in this figure are the same as for Figure \ref{0627_figs}.
The polar cap emission map in (d) assumes $r_{\rm em}$ = 45~km and in (e) assumes $r_{\rm em}$ = 230~km.
}
\end{figure*}

\begin{figure*}

\begin{minipage}{.49\textwidth}
\center
\includegraphics[width=7.5cm]{1828_z}
\subcaption*{}
\label{1828_z}
\end{minipage}
\begin{minipage}[t]{.49\textwidth}
\begin{minipage}[b]{.99\textwidth}
\begin{minipage}[b]{.45\textwidth}
\center
\includegraphics[height=4.2cm,angle=-90]{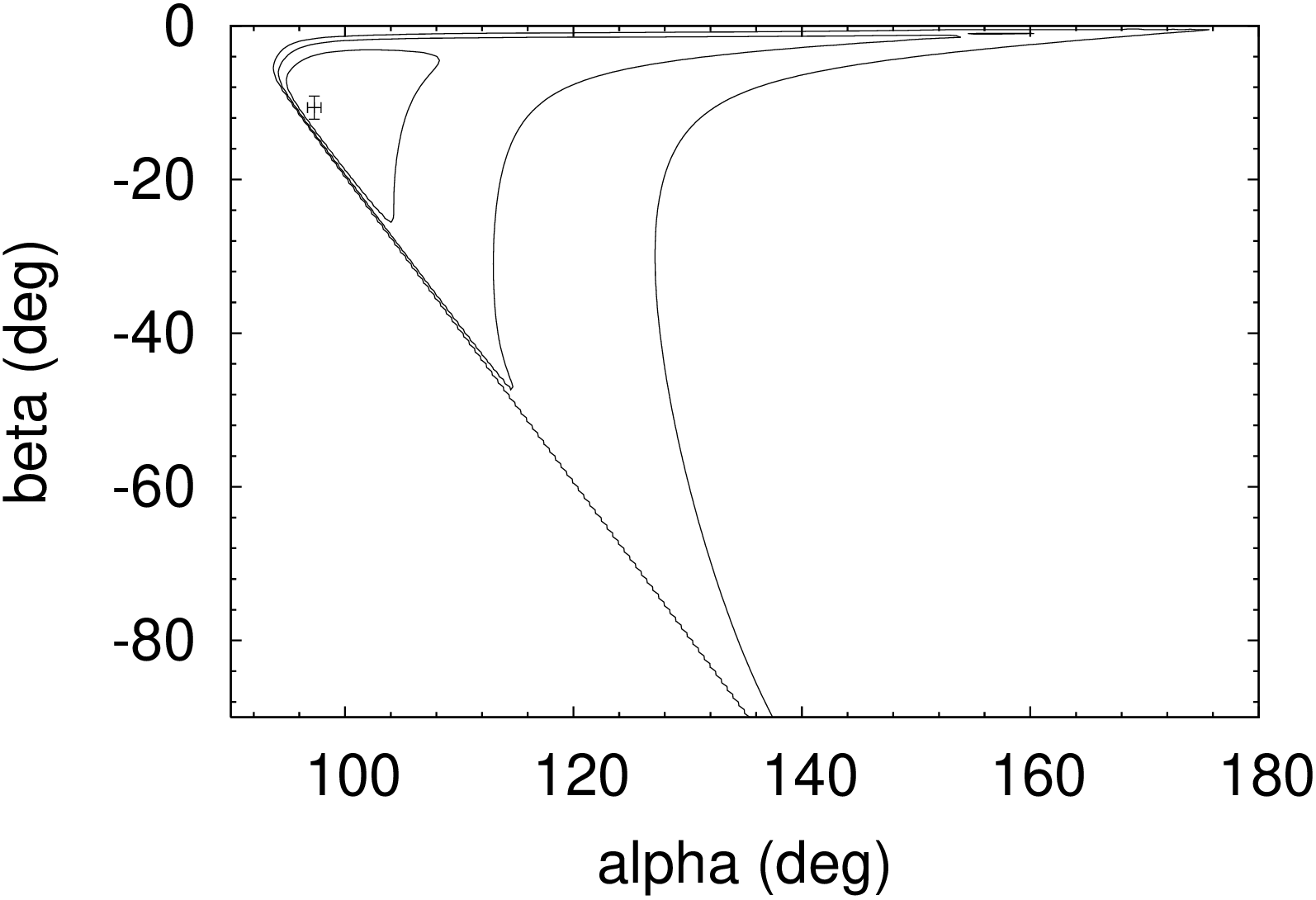}
\subcaption*{}
\label{1828_cont}
\end{minipage}
\begin{minipage}[b]{.45\textwidth}
\center
\includegraphics[height=5.5cm]{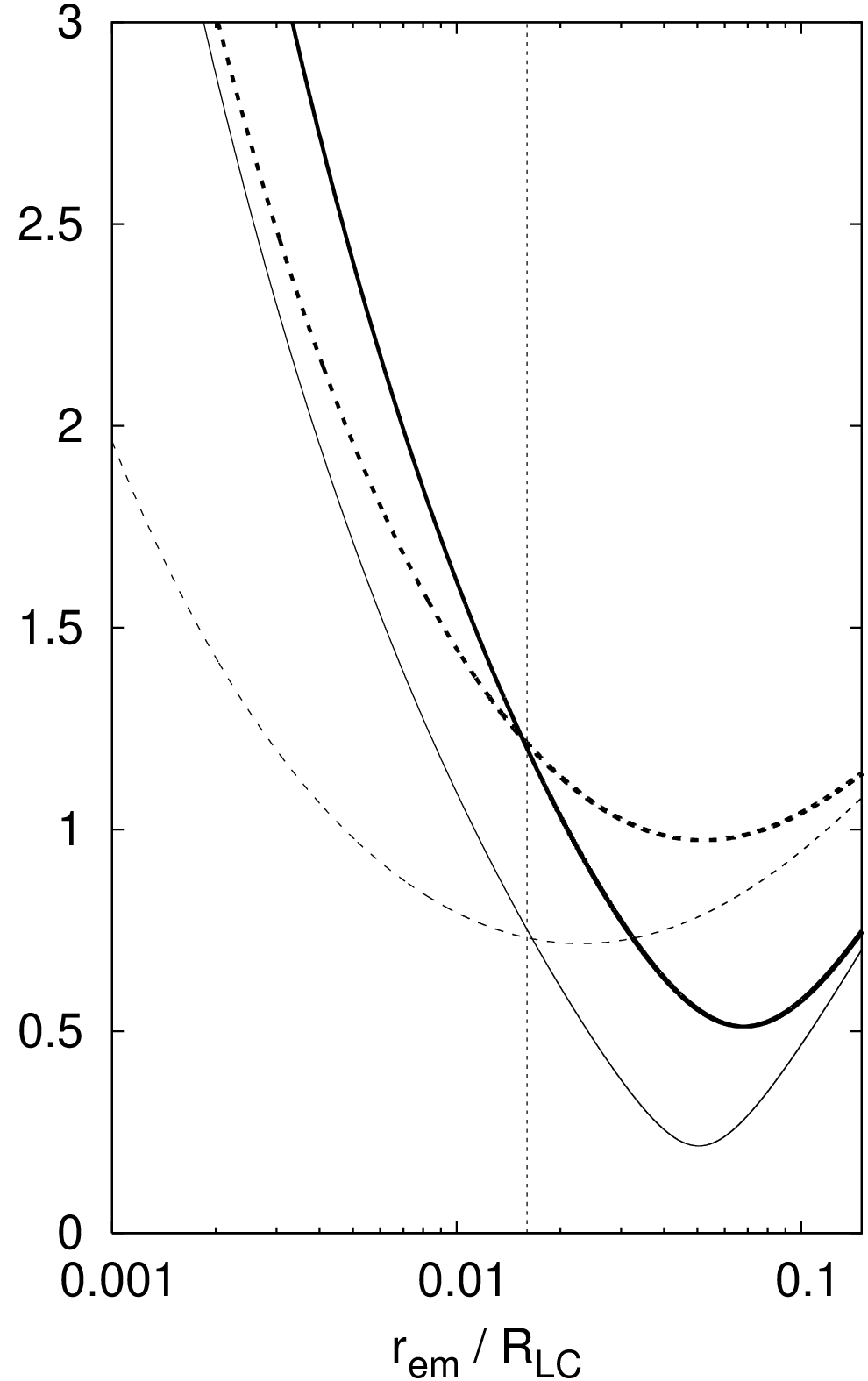}
\subcaption*{}
\label{1828_splot}
\end{minipage}
\end{minipage}
\begin{minipage}[b]{.99\textwidth}
\center
\includegraphics[height=3.8cm,angle=-90]{1828_beammap}
\subcaption*{}
\label{1828_beammap}
\end{minipage}
\end{minipage}
\caption[]{
\label{1828_figs}
(a) The 3.1~GHz integrated pulse profile of PSR~J1828--1101, clipped to show the details of the two pulses.
Note that there is an orthogonal mode jump between the MP and IP.
The format of this plot, and the others in this figure are the same as for Figure \ref{0627_figs}.
The polar cap emission map in (d) assumes $r_{\rm em}$ = 55~km.
}
\end{figure*}

\section{Emission height and polar maps}
\label{sec_polarmaps}
For emission which originates from the neutron star surface, $\phi_0$
corresponds to the point
where the magnetic axis passes closest to the line of sight.
If the emission height, $r_{\rm em}$, is small
with respect to the light-cylinder radius, $R_{\rm LC}$, then
the steepest gradient of the PA swing arrives later with 
respect to the corresponding total intensity emission.
The relative shift between the intensity of the emission and 
$\phi_0$ is given by
\begin{equation}
\Delta\phi = 4r_{\rm em}/R_{\rm LC}
\end{equation}
\cite{bcw91,ha01,dyk08}.
The phase offset, $\Delta\phi$, is often measured by assuming
a fiducial point on the profile which is either the profile peak or 
the symmetry centre (see e.g. \citealp{bcw91,jw06}).
This method has its difficulties
because the choice of fiducial point is somewhat subjective.
If one assumes the emission height is the same for both MP and IP,
then one expects that $\Delta\phi_{\rm PA}=180$\degr\
(see Table~\ref{rvm_table}). Table~\ref{rvm_table} also lists
$\Delta\phi_{\rm I}$, the separation
between the MP and IP in total intensity.
This is estimated by choosing an suitable fiducial point for both the MP 
and IP of each pulsar, nominally the centre of the pulse.

Recently, \citeN{ww09} have shown how to identify the active regions
of the polar cap given the observational data and an assumption
about the emission height. First they define a parameter, $s$,
the distance between the footpoint of the active field line and
the magnetic axis divided by the polar cap radius. The field lines
at the magnetic axis therefore have $s=0$ and the last open
field lines have $s=1$. The observational data consist of the
difference in longitude of the leading ($\phi_{\rm l}$) and trailing
($\phi_{\rm t}$) edges
of the profile with respect to $\phi_0$. These values for both the MP and IP 
for each pulsar are listed in columns 3 and 4 of Table~\ref{der_table}.
Generally, the location of $\phi_0$ is closer to the trailing edge 
than the leading edge.
This is expected from Equation~2, if the assumption is made that
the profile is intrinsically symmetric. Indeed, \citeN{bcw91} used
this assumption to derive emission heights in a range of pulsars.

Although the symmetry argument is appealing, it may not be correct,
especially if the emission is patchy across the beam \cite{lm88}.
\citeN{ww09} derived a formalism which allows $s$ to be computed as
a function of emission height and we use their methodology to produce
such a figure for each pulsar in our sample.
Then, by choosing an emission height we can produce a map
showing the emission regions above the polar cap.
What criteria should govern our choice of emission height?
We pick two possible examples.
The first enables us to obtain a symmetric solution, that is, one where 
the $s$ value is the same for the leading and trailing edges of the profile.
The second is to obtain a solution where $s<1$ so that the emission arises
from the open field lines of the polar cap. In the standard picture one
might expect both these conditions to be met simultaneously.
For two of the pulsars in our sample we find that this is not the case.
All the pulsars are considered in detail below.

\section{Discussion of individual pulsars}
\label{pulsars}
\subsection{PSR~J0627+0706 (Figure~\ref{0627_figs})}
The RVM fit shows that this pulsar is an orthogonal rotator.
The line of sight passes less than $\sim$1\degr\ from the magnetic 
axis corresponding to the IP emission,
with the MP passing further from the axis with $\beta\sim 9$\degr.
The peak of the MP occurs 5\degr\ earlier than $\phi_0$.

At 0.7 GHz the MP consists of steep rising edge followed by
a smoother trailing edge. At 1.4~GHz a second component can be
discerned in the trailing part of the profile which becomes more
prominent at 3.1~GHz. The MP has an overall width of $\sim$12\degr.
The LP is modest at
all frequencies and there is some negative circular polarisation (CP).
The PA swing is smooth with no sign of
orthogonal mode jumps.

The IP is broader than the MP with a width of some 20\degr, 
as might be expected given $|\beta_{\rm IP}| < |\beta_{\rm MP}|$.
The IP amplitude is around 20\%
of the MP and its spectral index is steeper. The profile
consists of a smooth rising and trailing edge and is double peaked,
with the two peaks separated by $\sim$4\degr. The LP
is rather low and there is some negative CP.
The PA swing is steep through the centre of the profile
and there is an orthogonal jump prior to the first pulse peak.

One interpretation of the profile is that the MP is a 
`partial profile' (e.g. \citealp{lm88})
where the leading edge of the beam is seen, but the trailing edge is absent.
As can be seen in Figure~\ref{0627_splot}, there are two choices of 
$r_{\rm em}$ that yield symmetric solutions.
The first solution with $r_{\rm em}=0.007 R_{\rm LC}=160$~km, (see Figure~\ref{0627_beammap1}) causes the IP to neatly fill the open field line region,
however the MP emission then occurs well outside the open field line region.
The second solution with $r_{\rm em}=0.014 R_{\rm LC}=320$~km, (see Figure~\ref{0627_beammap2}) places the MP symmetrically inside the open field line region, but now the IP is displaced towards the trailing edge of the beam.
There is no emission height which allows both the MP and IP to be 
symmetrically located.

\subsection{PSR J1549--4848 (Figure~\ref{1549_figs})}
The RVM fit shows the pulsar to be an orthogonal rotator with 
the line of sight cutting close to both magnetic axes.
For this pulsar it appears that the separation between the steepest 
gradients of the PA for MP and IP is
3\degr\ away from the expected 180\degr.
However, we argue that the similarity in width between the MP and IP,
combined with their similar $\beta$ values and the lack of an apparent 
shift in $\Delta\phi_{\rm I}$ strongly suggests that the emission height 
is the same for both the MP and IP, and the fitting is affected by unmodeled 
perturbations in the observed PA.

At 1.4~GHz the MP has a width of $\sim$20\degr and the profile
is lopsided, possibly consisting of 3 blended components.
The LP is moderate, reaching
a peak near the centre of the profile. The PA swing is
smooth and unbroken. At 0.7~GHz the leading component of the profile is
weaker than at 1.4~GHz whereas at 3.1~GHz it is stronger.
The RVM fit shows that the line of sight crosses the magnetic axis
very close to the peak of the profile.

The IP is somewhat narrower than the MP with a
width of some $\sim$15\degr\ at 1.4~GHz. Its amplitude is about
30\% that of the MP, this rises with increasing frequency and
so the IP has a shallower spectral index than the MP.
Its profile appear to be a triple structure with high LP
through the pulse centre. The PA traverse is broken by
an orthogonal jump between the central and trailing components.

Our choice of emission height, $0.013 R_{\rm LC}=180$~km, gives a 
symmetric solution, with the open field line region nearly filled 
for both the MP and IP (see Figure~\ref{1549_beammap}).

\subsection{PSR~J1722--3712 (Figure~\ref{1722_figs})}
The RVM fit shows that the pulsar is an almost perfectly orthogonal rotator,
with the line of sight cutting some 6\degr\ away from the magnetic
axis for both the MP and IP.
The peak of the profile leads $\phi_0$ by $\sim 5$\degr.

The MP has an overall width of $\sim$20\degr\ and the profile
shows some evolution with frequency but
is reasonably symmetrical over the range considered here. At 0.7~GHz
the profile has a steep trailing edge, this resolves into a small
trailing component at 1.4~GHz which is more prominent again at 3.1~GHz.
The LP fraction is $\sim$40\% at all three frequencies
and there is significant positive CP.
The PA swing is smooth and unbroken with no sign of
orthogonal mode jumps. 

The IP in this pulsar is rather weak, less than 10\% of the
MP amplitude. This makes it hard to discern any structure or
any trend with frequency but it appears to be double peaked
and virtually 100\% linearly polarised.

We choose an emission height of $0.026 R_{\rm LC}=290$~km which places 
the centroid of emission for both MP and IP coincident with the magnetic axis.
The polar cap map (Figure~\ref{1722_beammap}) shows that the open field 
line region is more-or-less filled with emission.

\subsection{PSR~J1739--2903 (Figure~\ref{1739_figs})}
The RVM fit constrains the magnetic and rotation axes to be within 5 to
10\degr\ of orthogonality.
The RVM fit shows that the magnetic axis crossing occurs almost
exactly in the profile centre.

The MP undergoes significant 
evolution with frequency.  The MP width at 0.7~GHz is $\sim$17\degr,
whereas at 1.4 and 3.1~GHz the width is more like $\sim$10\degr. 
It is possible that the low frequency profile is affected by scatter
broadening. At 1.4 and 3.1~GHz the profile is split into two
narrower, blended components. The trailing component is brighter than
the leading component at 1.4~GHz but the reverse is true at 3.1~GHz.
The LP and CP fractions are very low. There
is a hint that the CP changes from negative to
positive in the pulse centre at 1.4~GHz.

The IP is relatively strong in this pulsar, with an amplitude
of about 40\% of the MP at 0.7 and 1.4~GHz and 25\% at
3.1~GHz. The separation of the centroid of the MP and IP
is almost exactly 180\degr. 
The profile width is $\sim$15\degr\ at all frequencies.
At 1.4~GHz the profile consists of an asymmetric single component and
there is little evolution with frequency. The IP is relatively
highly linearly polarised and has negative CP.
The PA swing is unbroken across the pulse.

It is rather difficult to choose a sensible emission height for this pulsar.
There are no solutions for which the emission lies entirely within the 
open field line region.
From Figure~\ref{1739_splot}, we have chosen two emission heights, one 
with a symmetrical solution, and the second that forces the 
bulk of the emission to lie within the open field line region.
The first height, $r_{\rm em}=0.003 R_{\rm LC}=45$~km,
(see Figure~\ref{1739_beammap1}) shows the components symmetrically
located about the magnetic axis.
However the IP emission then comes from a location twice the 
radius of the polar cap, as does the outer edges of the MP.
The second solution, $r_{\rm em}=0.015 R_{\rm LC}=230$~km,
(see Figure~\ref{1739_beammap2}) is the best attempt at locating 
the emission within the open field line regions, however both 
MP and IP are significantly displaced towards the trailing edge of the beam.

\subsection{PSR~J1828--1101 (Figure~\ref{1828_figs})}
We find that no satisfactory RVM fit can be obtained without the addition 
of an orthogonal mode jump between the MP and IP and, in this case, the
RVM fit yields an orthogonal geometry.
The centroid of emission of the MP lags $\phi_0$ by approximately 2\degr.

PSR~J1828--1101 is highly scattered at 1.4~GHz and not detectable at
0.7~GHz. We therefore describe and show the 3.1~GHz profile.
The MP at 3.1~GHz consists of a single component with a total
width of $\sim$10\degr. The LP fraction is about
50\% and there is a smooth swing of PA across the pulse.
The CP is low.
The IP, which is about 20\% the amplitude of the MP 
lies almost exactly 180\degr\ from the MP. The profile consists of
a single component of width $\sim$14\degr\ and appears to be highly
linearly polarised with significant negative CP.
The PA swing is smooth across the profile.

We choose an emission height of $0.016 R_{\rm LC}=55$~km, which gives both
a symmetric solution for the MP and IP and ensures that $s<1$
(see Figure~\ref{1828_splot}).

\section{Discussion}
\label{discussion}
\subsection{Orthogonal rotator statistics}
If the $\alpha$ distribution in pulsars is random, then one would
expect $\sim$5\% of pulsars to have IP emission. The actual
fraction is lower than this: \citeN{wj08b} listed only 27 pulsars with
IP emission out of a total of nearly 1500 objects. The discovery of
PSR~J0627+0706 since then raises the number to 28.
Remarkably, at the time
of that paper, there was no firm evidence that {\it any} of these 27
pulsars had $\alpha\sim 90$\degr. There is a variety of explanations
for this: RVM fits are not possible in pulsars where the LP is very low,
where the IP is very weak, or where the PAs are distorted from their
geometrical expectations. The list also contains several `notorious' pulsars
(PSRs B0950+08, B1702--19, B1822--09 and B1929+10) where RVM fits
cannot distinguish between orthogonal and aligned rotators (see e.g. \citealp{ew01} and references therein).
PSRs B1702--19 and B1822--09 are particularly interesting because 
the single pulse data seem to show evidence that the MP and IP 
somehow communicate information \cite{wws07,gjk+94}.

We have convincingly shown that the five pulsars in this paper have
PA swings which yield $\alpha\approx 90$\degr.
In addition, the recent results on PSRs B0906--49 \cite{kj08} and
B1055--52 \cite{ww09} now demonstrate that at least seven pulsars are
orthogonal rotators. A single pulse study of these IP pulsars is warranted
in light of the data on PSRs B1702--19 and B1822--09.
The results for these pulsars strongly support
the `standard' model where the MP and IP are emission from 
corresponding poles of the pulsar but, in the next section consider the
possibility of inwards emission.

\subsection{Inward emission model}
\label{inwards}
An alternative to the standard view of observing emission from both 
poles is a model where emission is seen both outward and inward from an
emission zone above a single magnetic pole \cite{dzg05}.
In this model, the standard outward emission makes up one component of the profile and the inward emission, from the same pole, is seen $\sim 180$\degr\ later.
With a dipole field the inward emission originates from the same field lines that would be active in the forward emission from the other pole.
This makes it difficult to observationally distinguish the two cases.
Because the inward emission arises from the other side of the star, there is a delay in arrival which allows the separation of the peaks to exceed 180\degr.

If IP pulsars are associated with inwards emission from a single pole, we would expect that $\Delta\phi_{\rm PA} - 180^\circ = 180^\circ - \Delta\phi_{\rm I}$, which follows from the fact that the intensity and PA swing are shifted in opposite directions to the outwards case \cite{dzg05}.
For four of the pulsars in our sample, both $\Delta\phi_{\rm I}$ and
$\Delta\phi_{\rm PA}$ are close to 180\degr\ and no firm conclusions
can be drawn. However, for PSRs J0627+0706, J0908--4913 and B1055--52
the offset from 180\degr\ in the observed intensity cannot be 
explained by inwards emission and we believe this model can be ruled
out with some certainty.

\subsection{Active field line regions}
We have shown that the emission heights above both magnetic poles are rather 
similar for a given pulsar (except perhaps for PSR~J1549--4848).
If this is the case, one might expect that the profiles
should be wider for small values of $\beta$ than for large values.
However, the profile widths of the MP and IP are rather 
similar even when the difference in $\beta$ is large.
It is also clear that the observed profile luminosity also does 
not depend on $\beta$ in the way one would expect for a smoothly 
filled emission region.
Therefore these data support the view that the shape of the emission 
region in pulsars is complex, even when the observed profile appears simple.

Apart from PSR~B1055--52, the emission heights we derive are low, with values
of a few hundred km, or only $\sim$1\% of the light cylinder radius
(see Table~\ref{der_table}).
For PSRs J0908--4913, J1549--4848, J1722--3712 and J1828--1101 we obtain
a satisfactory solution to the emission height which yields symmetry of
the profiles around the magnetic axis and emission located well within
the open field line region. For PSR~J0627+0706 we find that either
the IP is symmetrically located about the magnetic axis
in which case the MP has emission from outside the open field lines,
or the MP lies within the open field lines and the IP is located on 
the far trailing edge of the polar cap.
For PSR~J1739--2903 the situation is more problematic.
In this case, a symmetrical solution gives emission from outside the open
field lines for both the MP and IP. Forcing the emission into
the open field lines means both MP and IP are significantly displaced
to the trailing part of the beam.
\citeN{ww09} found for PSR~B1055--52 that there is no choice of emission
height for which the emission comes from entirely within the open field 
line region.

\section{Conclusion}
\label{concs}
Analysis of the PA swings of five pulsars with IP emission shows that they can 
be well fitted with the RVM.
Although it is clear that there are small perturbations on top of the 
simple geometric model, the five pulsars discussed in this paper 
plus PSRs J0908--4913 \cite{kj08} and B1055--52 \cite{ww09} show convincing 
evidence that their magnetic and spin axis are close to orthogonal.

We use a technique developed in Weltevrede \& Wright (2009) in order to
map the emission regions the polar cap in each pulsar.
General conclusions which can be drawn are that the emission appears
to be patchy and does not conform to a uniform conal illumination
and that the emission height is similar at the two poles for a given pulsar.
We find that for three of our pulsars, emission maps can be made which
show both symmetry about the magnetic axis and emission confined to the
open field lines. In two cases, however, we find that either emission
arises from `closed' field lines or that the profiles are highly asymmetric
with respect to the magnetic pole.

\bibliographystyle{mnras}
\bibliography{journals,myrefs,modrefs,psrrefs,crossrefs}

\end{document}